\documentclass[twocolumn]{article} 
\usepackage[T1]{fontenc} 
\usepackage{fourier} 
\usepackage[english]{babel} 
\usepackage{graphicx}
\usepackage{chngcntr}
\usepackage{color}
\counterwithout{figure}{section}
\usepackage{amsmath}
\usepackage{amsfonts}
\usepackage{amssymb}
\usepackage{siunitx}
\usepackage{upgreek}
\usepackage{placeins}
\usepackage{float}
\usepackage{sectsty} 
\allsectionsfont{ \normalfont\scshape} 
\usepackage{caption}
\usepackage[export]{adjustbox}
\usepackage[affil-it]{authblk}
 
\usepackage{fancyhdr} 
\title{\huge  Nonexponential decay of a giant artificial atom}

\author[1]{Gustav Andersson}
\author[1,2]{Baladitya Suri}
\author[3]{Lingzhen Guo}
\author[1]{Thomas Aref}
\author[1]{Per Delsing}
\affil[1]{Department of Microtechnology and Nanoscience MC2, Chalmers University of Technology, Kemiv\"agen 9 SE-41296 G\"oteborg, Sweden}
\affil[2]{Department of Instrumentation and Applied Physics, Indian Institute of Science, Bengaluru 560012, India}
\affil[3]{Max Planck Institute for the Science of Light, Staudtstrasse 2, D-91058 Erlangen, Germany}
\date{\normalsize\today} 

\begin{document}

\maketitle 

\textbf{  
    In quantum optics, light-matter interaction has conventionally been studied using small atoms interacting with electromagnetic fields with wavelength several orders of magnitude larger than the atomic dimensions \cite{Brune1990,Raimond2001}. In contrast, here we experimentally demonstrate the vastly different \emph{giant atom} regime, where an artificial atom interacts with acoustic fields with wavelength several orders of magnitude \emph{smaller} than the atomic dimensions. This is achieved by coupling a superconducting qubit \cite{Schoelkopf2008} to surface acoustic waves at two points with separation on the order of 100 wavelengths. This approach is comparable to controlling the radiation of an atom by attaching it to an antenna. The slow velocity of sound leads to a significant internal time-delay for the field to propagate across the giant atom, giving rise to non-Markovian dynamics \cite{Guo2017}. We demonstrate the non-Markovian character of the giant atom in the frequency spectrum as well as nonexponential relaxation in the time domain.
    }
    
Cavity Quantum Electrodynamics (cavity QED) studies the interaction between atoms and photons in configurations where the atom is placed inside a cavity to enhance the interaction strength with the radiation field\cite{Goy1983,Miller2005}. In parallel, artificial atoms based on superconducting circuits have emerged in the last 20 years and over the past 15 years, superconducting circuit quantum electrodynamics has developed as an analogue to the cavity QED experiments\cite{Gu2017}. This has enabled probing the strong and ultrastrong coupling regimes of atom - light interaction to study exotic phenomena including vacuum Rabi splitting \cite{Wallraff2004} and the controlled generation of non-classical photon states \cite{Hofheinz2009,Wang2016}. A later development was waveguide QED where superconducting circuits were coupled to open transmission lines \cite{Hoi2011,Abdumalikov2010,Roy2017}. These experiments replace natural atoms and optical cavities with Josephson junction-based superconducting circuits behaving as artificial atoms strongly coupled to the field in open transmission lines or planar or 3D microwave resonators \cite{Paik2011,Roy2017}.

For atoms coupled to laser light in optical cavity QED and the interaction of microwaves with either Rydberg atoms in cavities or artificial atoms in circuit QED, treating the atom as a pointlike dipole is a valid approximation. In fact, in all the experiments mentioned above the size of the atom $L$ is at least an order of magnitude smaller than the wavelength $\lambda$ of the interacting radiation.

More recently, a superconducting transmon qubit \cite{Koch2007} was coupled to propagating surface acoustic waves (SAW) on a piezoelectric substrate \cite{Gustafsson2014}, demonstrating qubit decay by SAW emission and nonlinear reflection of SAW beams. Analogues of resonator QED architectures were realized in later experiments \cite{Manenti2017,Chu2017,Moores2017,Bolgar2017,Noguchi2017} where an artificial atom was coupled to the phonons inside an acoustic cavity \cite{Manenti2016}. Further work exploited superconducting qubits to generate and characterize phononic Fock states in SAW cavities as well as in bulk acoustic cavities \cite{Chu2017,Satzinger2018}. Owing to the slow propagation speed of sound in solids (3000 m/s), for a given frequency the wavelength of sound is five orders of magnitude smaller than that of light in vacuum. This property allow artificial atoms to interact with propagating acoustic fields beyond the small-atom approximation such that $L>\lambda$ \cite{Kockum2014}. Crucially to the experimental results presented here, the slow velocity of SAW further enables the engineering of time delays that are significant compared to the relaxation time as laid out in ref.~\cite{Guo2017} proposing the giant atom configuration we have implemented here. In this case the emitted excitation can be reabsorbed by the atom leading to non-Markovian effects.

Growing interest in quantum information science has spurred substantial theoretical investigations of non-Markovian open quantum systems, characterized by information back-flow from the environment \cite{Breuer2016}. In this context, the physical origin of the non-Markovianity is typically the coupling to a structured or finite reservoir. In contrast, we have realized a non-Markovian system by introducing deterministic time-delayed feedback through coupling a single quantum emitter at two distant points to a radiation field in one dimension. This feedback is intrinsic to the giant atom itself and can be engineered. In addition to enabling new parameter regimes in atom-field interactions, time-delayed feedback has potential applications in quantum information processing \cite{Pichler2016,Pichler2017}. It has been suggested that this form of non-Markovianity could be exploited to generate cluster states for universal measurement-based quantum computation requiring considerably less hardware resources than gate-based approaches \cite{Pichler2017}. Recent work \cite{Kockum2018} has also demonstrated that architectures involving multiple giant atoms can be designed to realize inter-atomic interactions while protecting the atoms from decohering into the waveguide, provided their coupling points overlap in a braided configuration. This makes giant atoms a relevant candidate for applications in quantum simulation.

The artificial atom in this experiment is a transmon qubit, consisting of a superconducting quantum interference device (SQUID) loop connected in parallel with an interdigital transducer (IDT) \cite{Gustafsson2014,Aref2015}. This interdigitated finger structure provides a shunt capacitance for the transmon as well as coupling to SAW at wavelengths matching the periodicity of the IDT, \SI{1.25}{\upmu m}. Due to the fixed periodicity of the finger structure, the coupling is frequency dependent and given by [supplementary]
\begin{equation}
    \gamma\left(\omega\right) = \frac{N_p K^2 \omega_{01}}{4}\left(\frac{\sin{X}}{X}\right)^2,
\end{equation}
where $\omega_{01}/2\pi$ is the qubit resonance frequency and $X= N_p \pi (\omega_\textup{IDT}-\omega)/\omega_\textup{IDT}$. This gives rise to a coupling strength that is proportional to the number of finger pairs $N_p$, with a coupling bandwidth that scales inversely with $N_p$.
Our devices are fabricated using aluminium on GaAs which has a SAW velocity of $v_{\textup{SAW}} = \SI{2900}{m/s}$. The piezo-electric coupling coefficient ($K^2 = 0.07\,\%$) of GaAs, while relatively low compared to materials such as lithium niobate and quartz \cite{Aref2015}, is still strong enough to ensure that the qubit lifetime $\tau$ is dominated by the acoustic coupling.

 The giant atom is realized by splitting the IDT into two electrically connected coupling points separated a by a distance $L$. Relevant sample parameters are listed in table~\ref{qubit params}. Due to the strong electromechanical coupling and slow propagation velocity of SAW, the propagation time $T=L/v_{\textup{SAW}}$ can be orders of magnitude larger than the excited state lifetime. In such a case, a phonon emitted into the transmission line at one coupling point of the qubit due to the spontaneous decay may later be reabsorbed by the qubit at the second point.  Considering the local IDT as a pointlike coupling center with coupling strength $\gamma$, the time-delay dynamics of the excited state amplitude $a_e$ in the absence of external driving is given by \cite{Guo2017}
\begin{equation}
\frac{\partial a_e\left(t\right)}{\partial t} = -i\omega_{01}a_e\left(t\right)-\gamma\left(a_e\left(t\right)+a_e\left(t-T\right)\right).
\label{eom}
\end{equation}

In the giant atom regime, characterised by $\gamma T \gtrsim 1$, the time delayed coupling term in eq.~(\ref{eom}) introduces markedly non-Markovian effects, which is the subject of this paper.

\begin{figure*}[h]
\begin{center}
\includegraphics[scale=0.12]{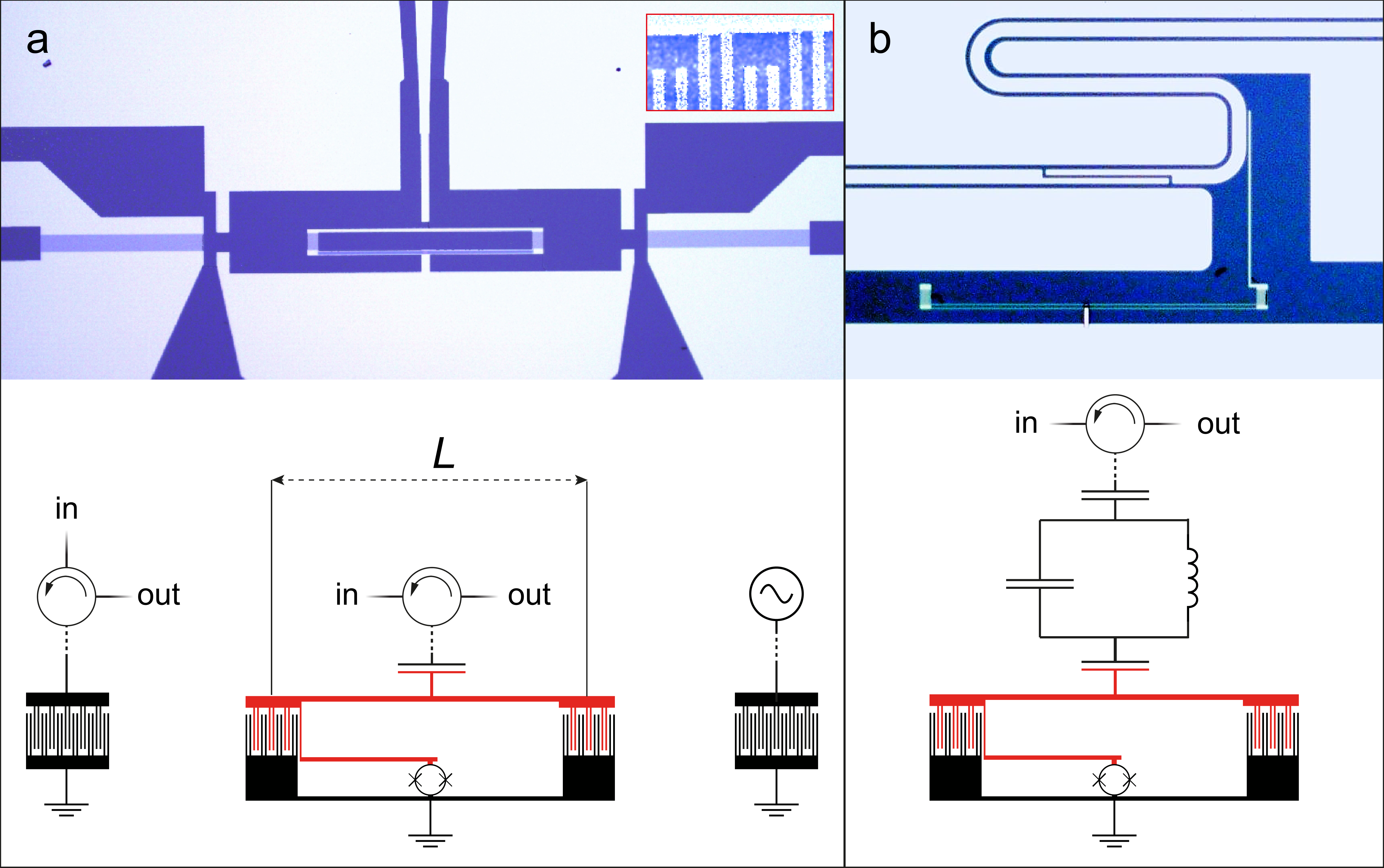}
\caption{\textbf{Device layouts.} \textbf{a}, false-color optical micrograph and circuit schematic of a giant atom capacitively coupled to an electrical gate. The GaAs substrate is shown in blue and the aluminum circuit and ground planes are white. The transmon circuit that makes up the giant atom is grounded and has two Josephson junctions forming a SQUID loop to enable frequency tuning via an external coil. The acoustic reflection and transmission properties are probed via IDTs on either side of the giant atom. A circulator routes the reflected gate signal to a cryogenic amplifier. Inset shows a false-color scanning electron micrograph of the finger structure of the giant atom IDTs. \textbf{b}, false-color optical micrograph and circuit schematic of giant atom coupled to a coplanar waveguide resonator. Excitation and readout tones are sent via a circulator to the resonator input. The reflected readout signal is amplified and measured similarly to the gate reflecion in \textbf{a}.}
\label{architectures}
\end{center}
\end{figure*}

The SAW-giant atom interaction is characterized using two different device designs. In design A, shown in Fig.~\ref{architectures}a, the transmon is embedded in a SAW transmission line with IDTs to generate and pick up acoustic signals. In addition to IDTs on either side, a capacitively coupled gate provides electrical control. Using this design we first measure the SAW scattering properties to verify acoustic coupling. The steady-state emission properties are then characterized for a range of giant atom parameters by measurements via the electrical gate. For a chosen set of giant atom parameters, we implement a second device, design B, where the transmon is capacitively coupled to a coplanar waveguide $\lambda/4$ resonator. This enables us to read-out the state of the transmon by probing the resonator, which we exploit to perform spectroscopy on the giant atom as well as time domain relaxation measurements. All the measured results are obtained at the \SI{10}{mK} base temperature of a dilution refrigerator. The qubit resonance frequency is tuned into the SAW coupling band ($\omega_{01}/2\pi\approx\SI{2.3}{GHz}$) using an applied external magnetic flux.
\begin{table*}
\begin{center}
\begin{tabular}{|c|c|c|c|c|c|c|c|}
\hline
sample & $N_p$ & $\gamma/2\pi$ & $T$ & $\gamma T$ & $L$ & $\gamma_\textup{gate}/2\pi$ & $2\gamma/\gamma_\textup{ext}$ \\ \hline
A1 & 14 & \SI{6.1}{MHz} & \SI{19}{ns} & 0.8 & \SI{55}{\upmu m} & \SI{1.25}{MHz} & 4.4\\
A2 & 14 & \SI{4.4}{MHz} & \SI{46}{ns} & 1.4 & \SI{125}{\upmu m} & \SI{1.5}{MHz} & 1.8\\
A3 & 18 & \SI{5.8}{MHz} & \SI{190}{ns} & 7.0 & \SI{550}{\upmu m} & \SI{2.2}{MHz} & 1.9\\
A4 & 18 & \SI{5.3}{MHz} & \SI{190}{ns} & 6.3 & \SI{550}{\upmu m} & - & - \\
B1 & 14 & \SI{4.8}{MHz} & \SI{160}{ns} & 4.8 & \SI{450}{\upmu m} & - & 1.4\\
\hline
\end{tabular}
\caption{\textbf{Giant atom sample parameters}. IDT finger pairs per coupling point $N_p$, Acoustic coupling strength per coupling point $\gamma$, time delay $T$ and coupling point separation $L$ for the giant atom samples. All samples have an IDT center frequency of \SI{2.3}{GHz}. The resonator in sample B1 has a resonance frequency of \SI{2.77}{GHz}. The coupling strength to the electrical gate $\gamma_\textup{gate}$ is shown for samples A1, A2, A3 in the second column from the right. The rightmost column shows the ratio of the total SAW coupling $2\gamma$ to the sum of non-acoustic decay rates $\gamma_\textup{ext}=\gamma_\textup{res}+
\gamma_\textup{gate}$. The parameters for samples A1, A2, A3 are extracted from fits to the gate reflection measurements shown in Fig.~\ref{scattering}b, where $\gamma$ is estimated by a fit of the resonance frequency to eq.~(\ref{eq. maxreflection}). The gate coupling $\gamma_\textup{gate}$ is estimated from the linewidth. The parameters for sample B1, which is coupled to a microwave resonator (with coupling strength $g=\SI{18}{MHz}$) rather than an electrical gate, are determined by two-tone spectroscopy of the absorption (Fig.~\ref{lineplots}a).}
\label{qubit params}
\end{center}
\end{table*}

\begin{figure*}
\begin{center}
\includegraphics[scale=0.66]{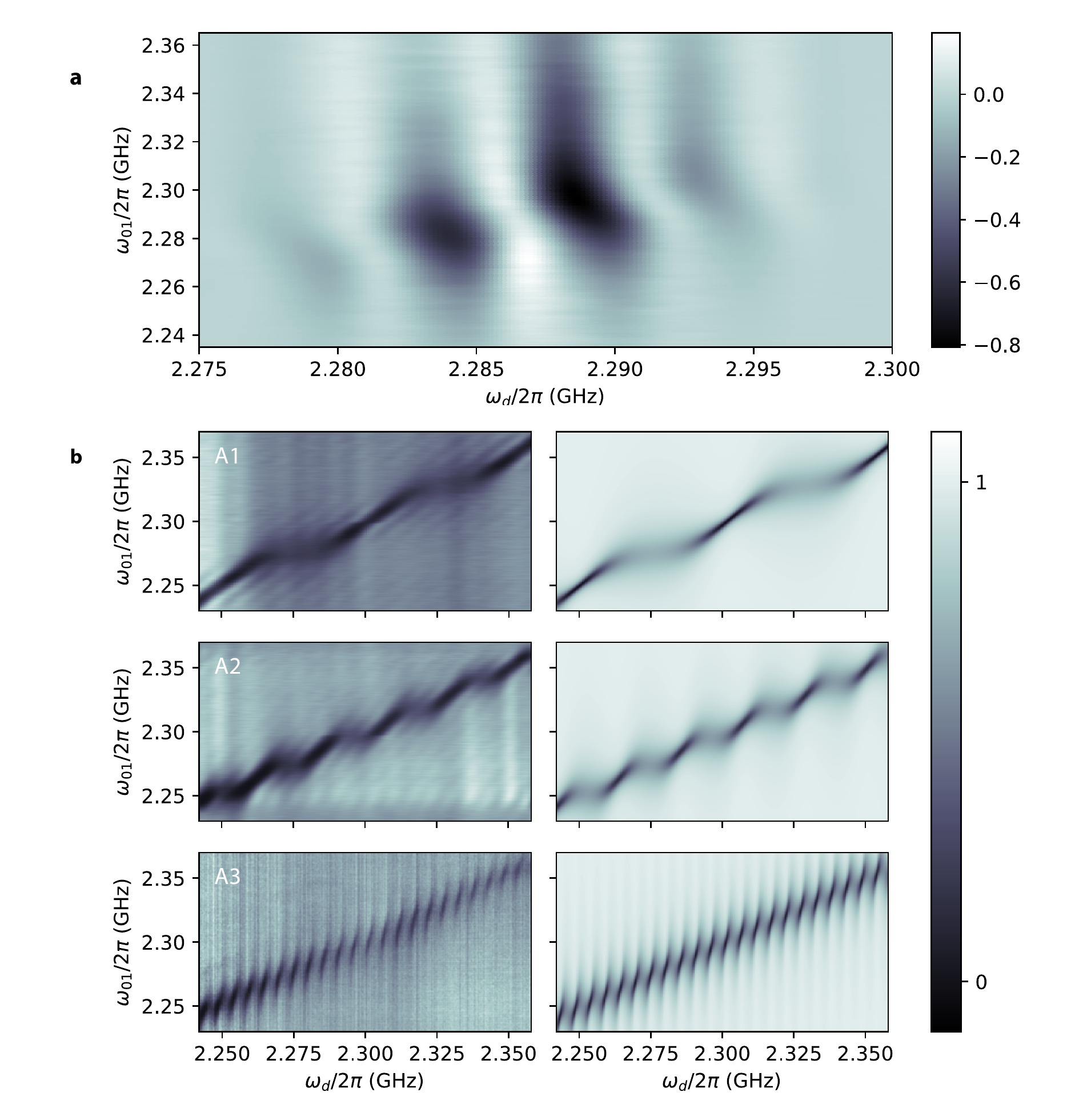}
\caption{\textbf{Scattering properties of the giant atom}. \textbf{a}, Acoustic power transmission (normalized) of a SAW signal from launcher IDT to pickup IDT measured as function of SAW frequency $\omega_d/2\pi$ and giant atom frequency $\omega_{01}/2\pi$ for sample A4. The background transmission when the giant atom is far detuned from the SAW frequency range has been subtracted. The small positive values for certain frequencies is an artefact of this subtraction [supplementary]. The magnitude is normalized to the maximal SAW transmission of the subtracted, far-detuned trace. The two visible dips in transmission are separated by $1/T \approx \SI{5}{MHz}$. The limited bandwidth of the launcher and pickup IDTs allows the acoustic scattering to be mapped across a frequency span of approximately \SI{14}{MHz}. \textbf{b}, Reflected electromagnetic power (normalized) off the capacitively coupled electrical gate for three different giant atoms. (A1-A3) The reflectance is measured while sweeping the drive frequency $\omega_d$ and giant atom frequency $\omega_{01}$ (left panel). Theoretical reflectance given by eq.~(8) of [supplementary] (right panel). Sample parameters are listed in table~\ref{qubit params}. The residual dissipation included to generate the theory plots are $\gamma_q/2\pi=\left \{\SI{1.5}{MHz},\SI{3.5}{MHz},\SI{4.0}{MHz} \right \}$ for A1, A2, A3. Increasing the coupling point separation, and thereby the time delay $T$, gives rise to finer structure in the interference pattern. The background reflection when the giant atom is far detuned from the SAW frequency range has been subtracted.}
\label{scattering}
\end{center}
\end{figure*}

The steady state acoustic transmission amplitude is measured at the pick-up IDT  while a continous weak SAW drive is applied to the giant atom using the launcher IDT. In Fig.~\ref{scattering}a we map the acoustic transmission amplitude for sample A4 as a function of drive frequency $\omega_d/2\pi$ while tuning the qubit resonance frequency $\omega_{01}/2\pi$. At weak driving powers, we expect to approach the limit of near-perfect extinction of transmitted SAWs when $\omega_d$ matches $\omega_{01}$  \cite{Astafiev2010, Hoi2011}. However, for our giant atom, we observe that interference between the scattering of each point causes the the condition for total reflection of a coherent beam to be modified from $\omega_{d} = \omega_{01}$. A right-propagating SAW beam of frequency $\omega/2\pi$ transmitted at $x=0$ (left coupling point) will pick up a phase factor $e^{i\omega T}$ before interacting with the giant atom again at $x=L$. The phase factors add up to yield a maximum reflection condition of \cite{Guo2017}

\begin{equation}
\omega_{01}=\omega_d-\gamma\sin{\omega_d T}.
\label{eq. maxreflection}
\end{equation}
The coupling point separation of sample A4 ($L=\SI{550}{\upmu m}$) is sufficiently large that interference fringes are visible within the bandwidth constrained by the launcher and pickup IDTs.

Next, we study the the SAW emission from the giant atom. A continuous drive of frequency $\omega_d/2\pi$ is applied to the capacitively coupled electrical gate while measuring the reflected amplitude of the drive tone. Efficient conversion of the input signal to SAW will result in a dip in the gate reflection, allowing for the characterization of the steady state emission of the giant atom (Fig.~\ref{scattering}b). Similarly to the acoustic reflection off a giant atom, this does not generally occur for $\omega_{01}=\omega_d$ [supplementary]. 



\begin{figure*}[!ht]
\begin{center}
\includegraphics[scale=0.75]{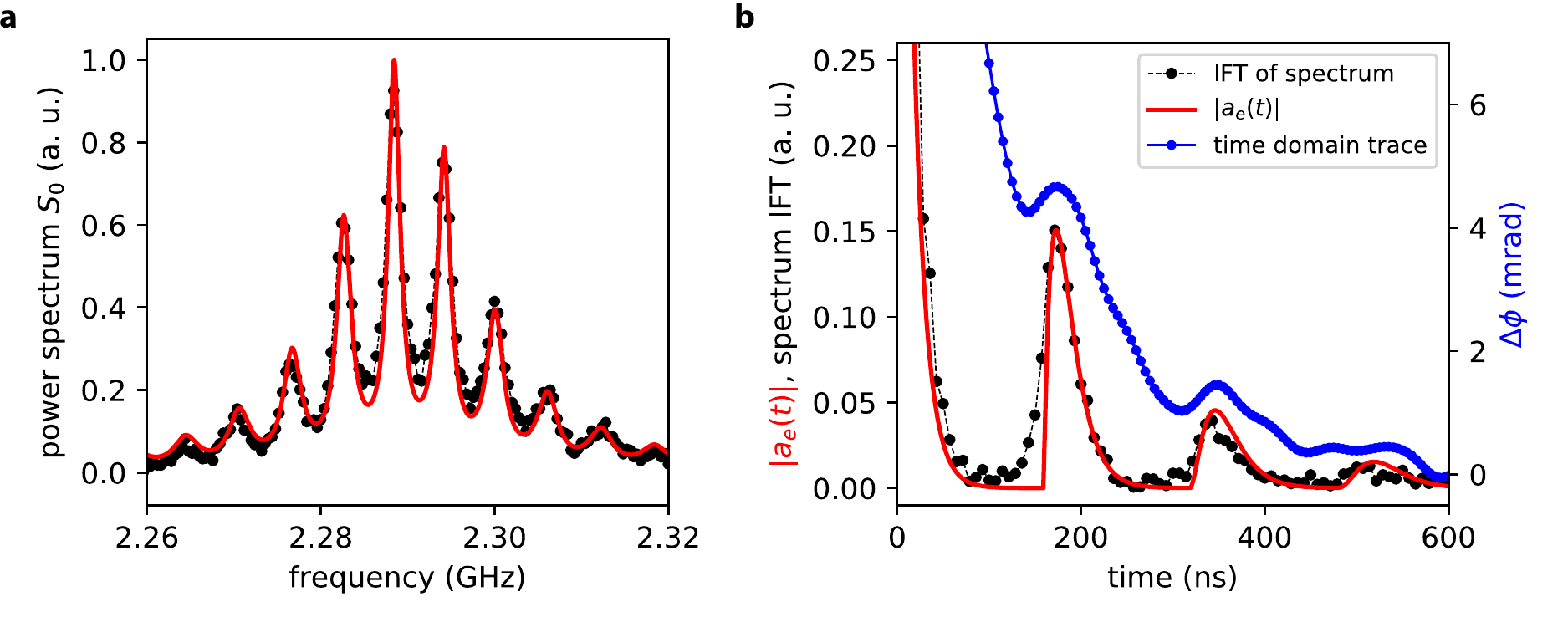}
\caption{\textbf{Frequency response and dynamics of the giant atom.} \textbf{a}, Two-tone spectroscopy of the giant atom measured through the electromagnetic resonator. A fixed-frequency readout is applied on resonance with the resonator, while a probe tone is swept across the giant atom resonance $\omega_{01}/2\pi = \SI{2.29}{GHz}$. The dotted black line shows the spectrum (normalized) of a giant atom with $L=\SI{450}{\upmu m}$ (sample B1) obtained from the readout phase response. A fit to eq.~(\ref{eq: spectrum}) is shown as a red line. The multi-peak structure arises due to the interference of SAW emission from the giant atom coupling points. \textbf{b}, Time evolution of the giant atom excited state population. The dotted black line shows the magnitude of the inverse Fourier transform of the (complex-valued) spectrum plotted in \textbf{a}. The red line shows the time evolution of $\left|a_e(t)\right|$ corresponding to the spectrum $S_0$ in \textbf{a}. This implies the (magnitude square) Fourier transform of $a_e$ gives the fit to $S_0$. The relaxation obtained in the time domain via a readout tone to the resonator is shown in blue. The elevated power in the readout tone causes slowdown in the qubit-resonator interaction, leading to the slowly decaying envelope in the response.}
\label{lineplots}
\end{center}
\end{figure*}

The condition for maximal transduction of the electrical signal to outgoing SAW coincides with maximal acoustic reflection and is given by eq.~(\ref{eq. maxreflection}). In addition, the giant atom coupling to SAW depends on the drive frequency as
\begin{equation}
\gamma_\textup{eff}=\gamma\left(1+\cos\omega_d T\right).
\end{equation}
As this measurement is not limited by the bandwidth of the launcher IDTs, the giant atom properties may be probed across a wider frequency range than in the purely acoustic case. We exploit this to measure gate reflection for three samples with intrinsic time delays of $T\approx \left \{ \SI{19}{ns}, \SI{45}{ns}, \SI{190}{ns} \right \}$. Going deeper into the giant atom regime, i.e. increasing $T$, leads to a finer interference structure in the gate reflection, which is shown in Fig.~\ref{scattering}b. Fitting the resonance frequency modulation to data from the largest atom, the designed separation distance (\SI{550}{\upmu m}) yields a SAW velocity of \SI{2906}{m/s} on GaAs, consistent with literature values at millikelvin temperatures \cite{Aref2015}.
Coupling the giant atom to an electromagnetic resonator in addition to the SAW field enables spectroscopy in the dispersive regime of circuit QED. This allows us to measure the excited state population. In sample B1, a $\lambda/4$ resonator with resonance frequency $f_r=\SI{2.77}{GHz}$ is capacitively coupled to the giant atom with coupling strength $g/2\pi=\SI{15}{MHz}$. Operating the giant atom at IDT resonance (\SI{2.29}{GHz}) gives an atom-cavity detuning of \SI{480}{MHz}, resulting in a dispersive shift $\chi/2\pi=\SI{0.47}{MHz}$. This is small compared to the SAW-atom interaction strength but sufficiently large to read out the qubit state via the resonator phase response. The spectrum of sample B1 is probed in two-tone spectroscopy with a weak, fixed-frequency readout tone applied at $f_r$ while a drive tone is swept across the resonance of the giant atom. The interaction with phonons already emitted into the SAW channel gives rise to a multi-peak structure in the absorption very different from the Lorentzian lineshape of ordinary (small) atoms. In Fig.~\ref{lineplots}a the phase of the readout tone is shown along with a fit to the expression
\begin{equation}
S_0\left(\omega\right)=\frac{\omega_{01}}{\left|\omega-\omega_{01}+i\gamma\left(1+e^{i\omega T}\right)+i\gamma_\textup{res}\right|^2}
\label{eq: spectrum}
\end{equation}
where $\gamma_\textup{res}\approx 2\pi \cdot\SI{6.85}{MHz}$ represents residual broadening of the spectral lineshape induced by non-acoustic decay as well as the finite spectroscopic drive power.

The spectrum of the giant atom excitation is related to its time evolution via the Fourier transform, such that $ S_0 = \left|\int_{-\infty}^{\infty}a_e(t) e^{i\omega t} dt \right|^2$. The excitation amplitude of an initially excited giant atom evolves according to
\begin{equation}
a_e\left(t\right)=\sum_{n=0}^\infty \Theta\left(t-nT\right)\frac{\left(-\gamma\left(t-nT\right)\right)^n}{n!}e^{-i\left(\omega_{01}-i\gamma -i\gamma_\textup{res} \right)\left(t-nT\right)}
\label{eq: a_e}
\end{equation}
where $\Theta\left(t\right)$ is the Heaviside function. For times $t>T$, interference between the instantaneous relaxation and the amplitude already emitted into the acoustic field gives rise to subexponential decay. While the inital decay is exponential with a rate fixed by $\gamma$, each revival peak decays slower than exponentially and, on sufficiently short timescales, the peak amplitude is only polynomially damped. The time evolution of $|a_e(t)|$ that corresponds to the power spectrum of Fig.~\ref{lineplots}a is shown as the red line in Fig.~\ref{lineplots}b. It captures well the features of the Inverse Fourier Transform (IFT) of the measured giant atom power spectrum (black dotted line, Fig.~\ref{lineplots}b). In the supplementary materials we show that the trace distance between the initially excited state and the ground state evolves non-monotonically in time, providing a quantitative measure of the non-Markovianity of the giant atom \cite{Breuer2009}.

To directly probe the non-exponential decay in the time domain, we monitor continuously the phase response of the readout resonator while applying an approximate $\pi$-pulse to the giant atom. The data is averaged approximately $10^7$ times and shown as the blue line in Fig.~\ref{lineplots}a. The readout tone is applied with high power \cite{Reed2010} such that the qubit-cavity system enters the nonlinear regime where switching between meta-stable states occur. In this regime the resonator acquires a memory of the qubit state longer than the qubit relaxation time, giving rise to a slowly decaying exponential envelope. This appears as a smearing effect in the time trace. Two revivals of the giant atom are clearly visible in this measurement, occurring at times consistent with the spectroscopic data.


Our results show that the acousto-electric interaction can be exploited to investigate previously unexplored parameter regimes of light-matter interaction. We have demonstrated that superconducting qubits can be strongly coupled to SAW such that the giant atom regime can be reached with significant internal time delays giving rise to non-Markovian dynamics. In the present experiment we have studied spontaneous emission from a single giant atom. Further new physics can be expected in circuit- and waveguide-QED settings for systems that break the Markov approximation. For example, it would be interesting to study waveguide-mediated interactions between quantum emitters in the non-Markovian regime, or the quantum properties of sound reflected from a giant atom.

\subsection*{Acknowledgements}
This work was supported by the Knut and Alice Wallenberg foundation and by the Swedish Research Council, VR. This project has also received funding from the European Union's Horizon 2020 research and innovation programme under grant agreement No 642688. We acknowledge fruitful discussion with M.K Ekstr\"om and G. Johansson.

\subsection*{Availability of data}
The data generated and analysed in this study are available from the corresponding author upon reasonable request.

\FloatBarrier

\bibliography{magabib}
\bibliographystyle{nature}

\end{document}